\documentclass[twocolumn,showpacs,aps,prl,raggedbottom,nobalancelastpage,amssymb,superscriptaddress]{revtex4}

\usepackage{graphicx}
\usepackage{amsmath}
\usepackage{amssymb}
\usepackage{bm}
\usepackage{color}

\newcommand{\ignore}[1]{}

\begin{document}
\title{
Symmetry causes a huge conductance peak in double quantum dots.}
\author{Robert S. Whitney}
\affiliation{
         Institut Laue-Langevin, 6 rue Jules Horowitz, B.P. 156,
         38042 Grenoble, France.}
\author{P. Marconcini} 
\author{M. Macucci}
\affiliation{
         Dipartimento di Ingegneria dell'Informazione, Universit\`a di Pisa, 
         Via Caruso 16, I-56122 Pisa, Italy. }

\date{\today}

\begin{abstract}
We predict a huge interference effect 
contributing to the conductance through large ultra-clean quantum dots of 
chaotic shape. When a double-dot structure 
is made such that the dots are the mirror-image 
of each other, constructive interference can make a tunnel barrier 
located on the symmetry axis effectively transparent. 
We show (via theoretical analysis and numerical simulation) that
this effect can be orders of magnitude larger 
than the well-known universal conductance fluctuations and weak-localization 
(both less than a conductance quantum).  A small magnetic field 
destroys the effect, massively reducing the double-dot 
conductance; thus a 
magnetic field detector is obtained, with a similar sensitivity to a SQUID, 
but requiring no superconductors.
\end{abstract}

\pacs{73.23.-b, 05.45.Mt,73.23.Ad,74.40.+k}

\maketitle

\begin{figure}
\includegraphics[width=6.5cm]{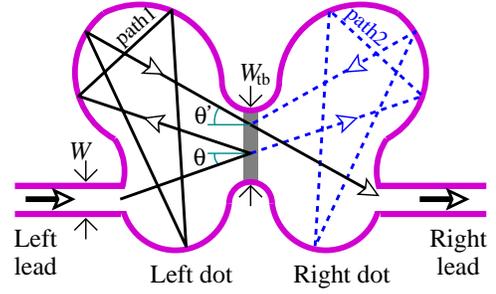}
\caption[]{\label{Fig:butter-path}
A mirror-symmetric double dot, where the classical dynamics
is highly chaotic.
We call it a ``butterfly double dot'' to emphasize the left-right symmetry.
Every classical path from the left lead to the right lead (solid line)
which hits the barrier more than once, is part of a family of 
paths which are related to it by the mirror symmetry (dashed line). 
}\end{figure}

\begin{figure}
\includegraphics[width=8.4cm]{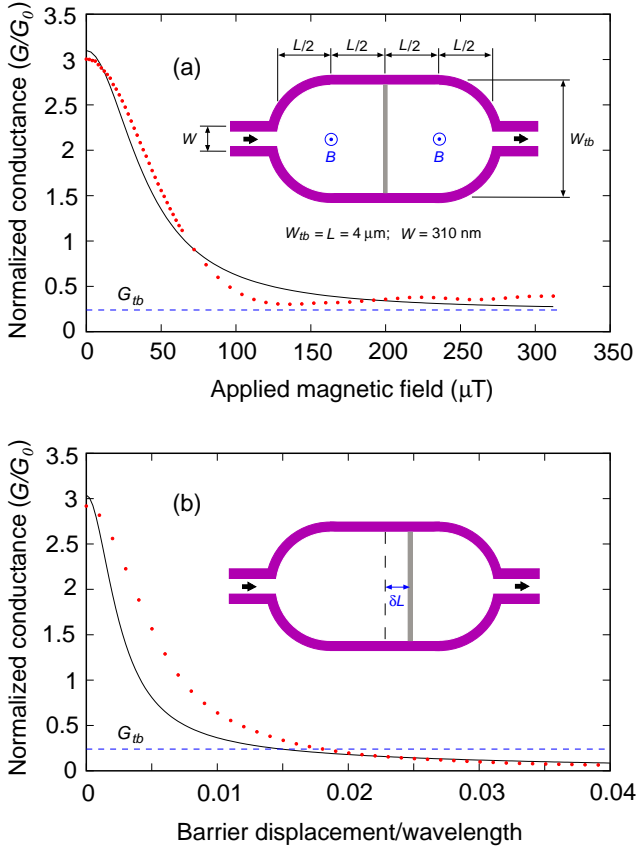}
\caption[]{\label{Fig:numerics}
Average conductance as (a) a function of applied $B$-field (with the barrier
on the symmetry axis),
and as (b) a function of the barrier position (for zero $B$-field).  
The latter mimics the effect of gates that reduce the size of one dot relative to the other.
The data points come from simulations performed for the structures 
shown in the insets.
The curve comes from the semiclassical theory; 
in (b) there is no fitting parameter, 
while in (a) an unknown parameter (or order one) is adjusted to 
fit the data. The
conductance of the
tunnel barrier alone is $G_{\rm tb}$.} 
\end{figure}

In the 1990s, interference effects (universal conductance fluctuations
and weak-localization) were observed for electrons flowing through 
clean quantum
dots \cite{Alhassid-review,Marcus-chaos}.
The chaotic shape of such dots makes these effects analogous to
speckle-patterns in optics rather than to the regular interference patterns
observed with Young's slits or Fabry-Perot etalons.
While such interference phenomena are beautiful, they have only a small
effect on the properties of quantum dots coupled to multi-mode leads.
Here we provide a theoretical analysis and numerical simulations showing that 
a much larger interference effect occurs in systems which are mirror-symmetric
but otherwise chaotic \cite{Baranger-Mello,Gopar-et-al,Kopp-Schomerus-Rotter,
Gopar-Rotter-Schomerus}, see Fig.~\ref{Fig:butter-path}.
We show that the mirror symmetry induces interference that 
greatly enhances tunneling
through a barrier located on the symmetry axis; it can make the
barrier become effectively transparent.
Thus an open double-dot system with an almost
opaque tunnel barrier between the two dots will exhibit a huge peak in
conductance when the two dots are the mirror image of each other,
see Fig.~\ref{Fig:numerics}.  This
effect could be used to detect anything which breaks the mirror symmetry.
For example, current 2D electron gas (2DEG) 
technology \cite{best-ultraclean-samples} could
be used to construct a device whose resistance changes by a factor of
ten, when an applied magnetic flux changes from zero to a fraction of a
flux quantum in the double dot.
This is a sensitivity similar to that of a SQUID, but it is achieved
without superconductivity, making it easy to integrate with other 2DEG
circuitry.

{\bf Origin of the conductance peak.}
The origin of the effect can be intuitively understood
by looking at Fig.~\ref{Fig:butter-path}.
Assume that electrons only follow the two paths shown 
(instead of an infinite number of different paths). 
Path 1 does not tunnel the first time it hits the barrier, 
but does tunnel the second time it hits it. Path 2 tunnels the first time it 
hits the barrier, but not the second time.  
Quantum mechanics gives the 
probability to go from the left lead
to the right lead as
$|r(\theta) t(\theta'){\rm e}^{{\rm i} S_1/\hbar} 
+ t(\theta) r(\theta'){\rm e}^{{\rm i} S_2/\hbar}|^2$, where 
the scattering matrix of the tunnel barrier has amplitudes
$r(\theta)$ and $t(\theta)$
for reflection and transmission at angle $\theta$.
If there is no correlation between the classical actions of the two paths
($S_1$ and $S_2$), then the cross-term cancels upon averaging over energy,
leaving the probability as 
$|r(\theta) t(\theta')|^2+|t(\theta) r(\theta')|^2$.
In contrast, if there is a perfect mirror symmetry, then $S_2=S_1$,
and the probability is 
$|r(\theta) t(\theta')+t(\theta) r(\theta')|^2$, 
which is significantly greater than 
$|r(\theta) t(\theta')|^2+|t(\theta) r(\theta')|^2$.
Indeed, if we could drop the $\theta$-dependence of 
$r$ and $t$,
the probability would be doubled by
the constructive interference induced by the mirror symmetry.
A path that hits the barrier
$(n+1)$ times has $2^n$ partners with the same classical action
(each path segment that begins and ends on the barrier can be reflected
with respect to the barrier axis). 
However the conductance is {\it not} thereby enhanced by $2^n$,
because (due to the nature of the barrier scattering matrix)
there is also destructive interference
when one path tunnels $(4j-2)$ times more than another
(for integer $j$).

The effect looks superficially like resonant tunneling.
However, that only occurs when dots are weakly 
coupled to the leads, so that each dot
has a peak for each level of the closed dot
and the current flow is enhanced when two peaks are aligned.   
Instead in our case
each dot is well coupled to a lead
(with $N\!\gg\!1$ modes), so the density of states
in each dot is featureless 
(the broadenning of each level is about $N$ times
the level-spacing). Furthermore, resonant tunneling occurs
at discrete energies, while our effect is largely energy independent.
Another similar effect, called ``reflectionless tunneling'',
occurs when electrons are 
{\it retro-reflected} as holes, due to Andreev reflection from
a superconductor
\cite{reflectionless-tunnel-expt91,
reflectionless-tunnel-review}.  
However, this retro-reflection
transforms the classical dynamics in the dot from chaotic to integrable
\cite{Kosztin-Maslov-Goldbart}, 
and large interference effects in
integrable systems are not uncommon (consider a Fabry-Perot etalon).
Here, the mirror symmetry induces a large interference effect 
without any retro-reflection
and without a change in the nature of the classical dynamics
(chaotic motion remains chaotic).

\begin{figure}
\includegraphics[width=6.6cm]{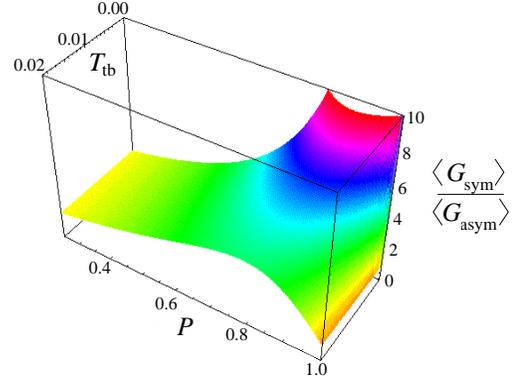}
\caption[]{\label{Fig:cond-ratio}
Plot of the ratio $\langle G_{\rm sym}\rangle/\langle G_{\rm asym}\rangle$,
given by Eqs.~(\ref{eq:Gsym},\ref{eq:Gasym}).
The ratio grows as $T_{\rm tb}\to 0$ for all $P$
(although $\langle G_{\rm sym,asym}\rangle$ shrink).
For given $T_{\rm tb}$, the ratio is maximal
at $P=(1 -2T_{\rm tb}^{1/2})/(1-4T_{\rm tb})$.
} 
\end{figure}


{\bf Semiclassical theory.}
Our analysis uses the semiclassical theory
of transport through clean chaotic quantum dots \cite{Bar93}.
The conductance through a system whose dimensions are much greater than a 
Fermi wavelength can be written as a double sum over
classical paths, $\gamma$ and $\gamma'$, which both start at a point $y_0$
on the cross-section of 
the left lead and end at $y$ on the right lead:
\begin{eqnarray}
G
&=&  
(2\pi \hbar)^{-1} G_0\sum_{\gamma,\gamma'} 
A_{\gamma}A_{\gamma'}^*
\exp \big[{\rm i}(S_\gamma-S_{\gamma'})/\hbar \big] ,
\label{eq:conductance}
\end{eqnarray}
where $G_0= 2e^2/h$ is the quantum of conductance, 
and $S_\gamma$ is the classical action of path $\gamma$.
A tunnel barrier with left-right symmetry must have the scattering matrix
\begin{eqnarray}
{\cal S}_{\rm tb}(\theta) = {\rm e}^{{\rm i} \phi_{r(\theta)}} 
\left(\begin{array}{cc} |r(\theta)| & \pm {\rm i} |t(\theta)| \\
\pm {\rm i} |t(\theta)| & |r(\theta)|\end{array} \right)  
\label{eq:Stb}
\end{eqnarray}
where $r(\theta)$ and $t(\theta)$ are reflection and transmission amplitudes 
for a plane wave at angle of incidence $\theta$.    
Keeping only the upper sign in ${\cal S}_{\rm tb}(\theta)$ 
\cite{footnote:sign},  
the amplitudes in Eq.~(\ref{eq:conductance})
are
\begin{eqnarray}
\label{eq:A}
A_\gamma &=&  
\left(\frac{{\rm d} p_{y_0}}{{\rm d} y}\right)^{1/2}_{\gamma}\   
\prod_{j=1}^{m_{\rm T}(\gamma)}  {\rm i}|t(\theta_{{\rm T}j})| 
\prod_{k=1}^{m_{\rm R}(\gamma)} |r(\theta_{{\rm R}k})| \ \ 
\end{eqnarray}
where path $\gamma$ starts with a momentum across the left lead, 
$p_{y_0}$, and a total
momentum given by the Fermi momentum, $p_{\rm F}$.
This path 
reflects off the barrier $m_{\rm R}(\gamma)$ times
(with the $k$th reflection at angle $\theta_{{\rm R}k}$)
and transmits $m_{\rm T}(\gamma)$ 
times (at angles $\theta_{{\rm T}k}$)
before hitting the right lead at $y$. 
The factor $({\rm d} p_{y_0}/{\rm d} y)_\gamma$
is the stability of the path that would exist if the
barrier were absent for each transmission and impenetrable for each
reflection. 
For most pairs with $\gamma\neq\gamma'$, the exponent in 
Eq.~(\ref{eq:conductance}) varies fast with energy,
so that averaging over energy removes such pairs from the double sum.
We keep only the main contributions surviving such averaging:
those where $\gamma'$ can be constructed from $\gamma$ by means of the 
reflection with respect to the barrier axis (symmetry axis)
of any path segment that begins and ends on the barrier,
for which $S_{\gamma'}=S_\gamma$ at all energies
(the paths thereby have the same stability 
$({\rm d} p_{y_0}/{\rm d} y)_\gamma$).
Dropping weak-localization effects
\cite{Richter-Sieber,Baranger-Mello}, 
the average conductance reads
\begin{eqnarray}
\langle G \rangle 
&=& {\frac{G_0} {2\pi \hbar}}
\!\int_{\rm L} \! \! {\rm d} y_0 \int_{\rm R} \! {\rm d} y   
\sum_\gamma
\left|{\frac{{\rm d} p_{y_0}} {{\rm d} y}}\right|_\gamma
\left[{\prod_{m=1}^{n(\gamma)}} {\mathbb C}_{\gamma,m} 
{\mathbb S} \right]_{41} \ \
\label{eq:conductance-diag}
\end{eqnarray}
where the product is ordered, and $n(\gamma)$ is the number of times 
the path $\gamma$ hits the barrier.
The four-by-four matrix ${\mathbb S} = {\cal S}_{\rm tb} \otimes {\cal S}_{\rm tb}^\dagger$
gives the scattering of the two paths at the barrier.
Thus ${\mathbb S}_{ij}$ gives the weight to go from state $j$ to state $i$,
where we define state 1 as both paths in the left dot;
state 2 as path $\gamma$ in the left dot and path $\gamma'$ in the right dot;
state 3 as path $\gamma$ in the right dot and path $\gamma'$ in the left dot;
and state 4 as both paths in the right dot.
The matrices ${\mathbb C}_{\gamma,m}$ 
are diagonal with the following non-zero elements: 
$[{\mathbb C}_{\gamma,m}]_{11}=[{\mathbb C}_{\gamma,m}]_{44}=1$
and 
$[{\mathbb C}_{\gamma,m}]_{22}=[{\mathbb C}_{\gamma,m}]_{33}^*
=\exp[{\rm i} \delta S_{\gamma,m}/\hbar]$.
The action difference $\delta S_{\gamma,m}$ is that between path $\gamma$ 
in the left dot
and its mirror image in the right dot between the
$(m-1)$th and $m$th collision with the barrier.  
For perfect symmetry ${\mathbb C}_m = {\mathbb I}$ and then the product equals
$[{\mathbb S}^n]_{41}$.

We assume that the classical dynamics is sufficiently
mixing that paths uniformly explore the dot between 
subsequent collisions with the barrier (or leads).
Defining $\delta S_0/\hbar$ as the phase difference acquired in one time of flight
across the dot, we have  
${\mathbb C}_{\gamma,m} \simeq \exp[-\Gamma t_{\gamma,m}]$
where $\Gamma$ is a complex number,
with ${\rm Im}[\Gamma]\simeq \langle \delta S_0 \rangle/(\tau_0\hbar)$
and ${\rm Re}[\Gamma]\simeq {\rm var}[\delta S_0]/(\tau_0\hbar^2)$.
The probability that a path survives in the dot
for a time $t$ without hitting either the barrier or the lead is 
${\rm e}^{-t/\tau'_{\rm D}}$. 
Using this, we replace 
${\mathbb C}_{\gamma,m}$ by its time-average 
${\mathbb C} = \langle {\mathbb C}_{\gamma,m}\rangle$;
its only non-zero elements are
${\mathbb C}_{11} ={\mathbb C}_{44}=1$
and 
${\mathbb C}_{22} ={\mathbb C}_{33}^*= [1+\Gamma \tau'_{\rm D}]^{-1}$.
Thus the product in Eq.~(\ref{eq:conductance-diag})
reduces to $({\mathbb C} {\mathbb S})^n$.
The sum is over all $\gamma$s 
that hit the barrier $n$ times, and is independent of $y_0,y$. 
To proceed, we define 
$\tilde{\mathbb S} \equiv {\mathbb C}^{1/2}{\mathbb S} {\mathbb C}^{1/2}$;
it is simple to show that 
$\big[({\mathbb C}{\mathbb S})^n\big]_{41}=
[\tilde{\mathbb S}^n]_{41}$ for all $n$.
Then, 
defining $P= W_{\rm tb}/(W_{\rm tb}+W)$ as the probability for 
a path to hit the $W_{\rm tb}$-wide barrier before 
escaping into the $W$-wide lead, we find that
$\langle G \rangle = 
G_0 N (1-P) \sum_{n=1}^\infty P^n \big[\tilde{\mathbb S}^n \big]_{41}$,
where $N=p_{\rm F}W/( \pi \hbar)$ is the number of modes in a lead.
Upon finding the matrix, ${\mathbb U}$, which diagonalizes 
$\tilde{\mathbb S}$,
one can easily evaluate the geometric series in $n$.

This analysis gives the following average conductance of the
symmetric double dot ($\Gamma=0$),
\begin{eqnarray}
\langle G_{\rm sym} \rangle  &=& G_0N P (1+P)T_{\rm tb}/ 
[(1-P)^2 + 4PT_{\rm tb}],\ 
\label{eq:Gsym}
\end{eqnarray}
where $T_{\rm tb}$ is the tunneling probability, 
$|t(\theta)|^2$,
averaged over all $\theta$.
For $T_{\rm tb} < (1-P)/2$ (i.e. for $G_{\rm tb}$, 
the conductance of a barrier with transmission 
$T_{\rm tb}$ in a waveguide of width $W_{\rm tb}$,
less than
P times the conductance of the series of the two constrictions), one finds 
that $\langle G_{\rm sym}\rangle$ is greater (often much greater)
than the tunnel barrier conductance, $G_{\rm tb}$.
Thus symmetrically placing constrictions on either side of the barrier
can strongly {\it enhance} its conductance 
(this is a stark example of the fact that quantum 
conductances in series are not additive). 
In contrast, for the
asymmetric double dot (large $\Gamma$) we have
\begin{eqnarray}
\langle G_{\rm asym} \rangle &=& 
G_0N PT_{\rm tb}/[1-P+2PT_{\rm tb}],
\label{eq:Gasym}
\end{eqnarray}
which is always less than $G_{\rm tb}$.
The ratio
$\langle G_{\rm sym} \rangle/\langle G_{\rm asym} \rangle$
is plotted in Fig.~\ref{Fig:cond-ratio}.
For any finite $T_{\rm tb}$, the ratio is maximal
at $P=(1 -2T_{\rm tb}^{1/2})/(1-4T_{\rm tb})$.
This choice of $P$ gives $\langle G_{\rm sym} \rangle= G_0 N/4$
and (for small $T_{\rm tb}$) 
$\langle G_{\rm asym} \rangle\simeq T_{\rm tb}^{1/2}  G_0 N/2$.
Thus the conductance ratio can be arbitrarily large
for a highly opaque tunnel barrier.

{\bf Peak shape with symmetry-breaking.}
The effect of the mirror symmetry 
is suppressed by a perpendicular magnetic field, $B$,
or by moving the boundary of one dot
by a distance $\delta L$. It is also suppressed
by disorder (defined by a mean free flight time between 
subsequent scatters
from disorder, $\tau_{\rm mf}$) or decoherence
(defined by a decoherence time, $\tau_\varphi$).
The suppression can be quantified in terms of
the following parameters:
\begin{eqnarray}
\Gamma_B &=& \eta(eB {\cal A}/h)^2/\tau_0 ,
\label{eq:Gamma_B}
\\
\Gamma_{\rm boundary} &=&  \tau_0^{-1} \big({\rm var}[\delta L]/\lambda_{\rm F}^2
+{\rm i} \langle \delta L \rangle/\lambda_{\rm F}\big) , 
\\
\Gamma_{\rm mf} &=& \tau_{\rm mf}^{-1} ,
\qquad 
\Gamma_{\rm \varphi} = \tau_{\varphi}^{-1} ,
\end{eqnarray}
where $e$ is the electronic charge,
${\cal A}$ is the area of one dot, and $\tau_0$ is the time 
to cross the dot. 
In $\Gamma_B$,  the constant $\eta$ is of order one, 
but is hard to estimate \cite{footnote:kappa}.
For $\Gamma_{\rm boundary}$, we have 
$\langle\delta L\rangle \sim x\xi$ and 
${\rm var}[\delta L] \sim x^2(\xi-\xi^2)$,
if a fraction $\xi$ of the left dot is deformed outwards by 
a distance $x$.
For multiple asymmetries, 
the total $\Gamma$ is
the sum of the individual $\Gamma$s given above.
For real $\Gamma$,  
\begin{eqnarray}
\langle G(\Gamma) \rangle = \langle G_{\rm asym} \rangle + {\frac
{\langle G_{\rm sym} \rangle - \langle G_{\rm asym} \rangle}  
{1+ F(P,T_{\rm tb}) \times\Gamma\tau'_{\rm D}}},
\label{eq:peak-shape-realGamma}
\end{eqnarray}
where 
$F(P,T_{\rm tb})
=\langle G_{\rm sym} \rangle/[\langle G_{\rm asym} \rangle(1+P)] $,
and $\tau'_{\rm D} \sim \pi L\tau_0/(W+W_{\rm tb})$ 
is the typical time a path spends in one dot before
hitting a lead or the barrier.
For the large conductance ratio (see below Eq.~(\ref{eq:Gasym})),
$F(P,T_{\rm tb})\tau'_{\rm D}$ is about half the 
dwell time in the double-dot, 
$\tau_{\rm D}\sim (1-P)^{-1}\tau'_{\rm D}$.
Thus the conductance is a Lorentzian function of the $B$-field,
with similar width to the weak-localization dip in 
the same system with no barrier \cite{Richter-Sieber}.
This makes the system an extremely sensitive detector of magnetic fields
and deformations of the confining potential (for example
due to the movement of charge near the double dot).
Intriguingly, the peak remains when the 
leads are at different positions on the two dots;
it is simply suppressed with an asymmetry parameter
$\Gamma_{\rm lead} =(1-P)/\tau'_{\rm D}$.

For complex $\Gamma$, as in Fig.~\ref{Fig:numerics}(b), 
we have no analytic result for 
$\langle G(\Gamma) \rangle$, 
but we can get it by numerically 
diagonalizing the 4-by-4 matrix, $\tilde{\mathbb S}$. 
In Fig.~\ref{Fig:numerics}(b), the data and the theory curve 
drop below  $\langle G_{\rm asym}\rangle=0.23G_0$.
We will show elsewhere
that this is due to destructive interference.
The conductance rises back up to $\langle G_{\rm asym}\rangle$
when the barrier is moved a distance of order a wavelength.

{\bf Proposal for experimental observation.} 
Consider making such a double-dot in an 
ultra-clean two-dimensional electron gas (2DEG) at the lowest 
achievable temperatures \cite{best-ultraclean-samples}.
A finger gate could define the barrier 
\cite{barrier-finger-gate},
with split gates controlling the lead widths.
To maximize the effect for a 2DEG with a mean free 
path~\cite{best-ultraclean-samples} of order $500\,\mu$m,
each dot (see Fig. 2) can have size $L=\,4\,\mu$m
(circumference $\sim 3.6 L\sim 15\,\mu$m) with 12 mode leads
($W = 310\,{\rm nm} \sim 6\lambda_{\rm F}$). 
A barrier with $T_{\rm tb}= 1.48 \times 10^{-3}$ and width $W_{\rm tb} = L$
gives $P = 0.93$ and $\tau'_{\rm D} \sim 3.5\tau_0$.
In this case, $\langle G_{\rm sym} \rangle \simeq
14\langle G_{\rm asym} \rangle \simeq
3.2 G_0$ (resistance $R_{\rm sym} \sim 5\,{\rm k}\Omega$).
The crossover from
$\langle G_{\rm sym} \rangle$ to 
$\langle G_{\rm asym} \rangle$ happens for $\Gamma \simeq 0.14/\tau'_{\rm D}
\sim 0.04/\tau_0$. 
At low temperatures ($\tau_{\varphi} > \tau_{\rm mf}$), 
disorder will suppress the peak to about 83\% of 
$\langle G_{\rm sym} \rangle$,
since $F(P,T_{\rm tb}) \Gamma_{\rm mf}\tau'_{\rm D} \sim 0.2$.
Thus the double-dot conductance will
drop by an order of magnitude if $10\%$ of the boundary of one dot is
moved by $\lambda_{\rm F}/2$, or
if a B-field is applied such that a fifth of a flux-quantum 
threads each dot. The latter is a B-field sensitivity similar 
to that of a SQUID.

The main experimental challenge will be to define dots
that are mirror-symmetric on a scale significantly less than
$\lambda_{\rm F} \sim 50\,{\rm nm}$.
We suggest that 
each dot should be defined by means of multiple gates
(made as symmetric as possible); their 
voltages can then be tuned to maximise the symmetry. 
We propose the following protocol for this maximization.
Starting with very wide leads, in such a way that
$P$ is far from unity and the conductance peak is very broad,
one scans the
dot-defining gate voltages over a broad range to reveal the 
approximate symmetry point (maximal conductance).  
One then narrows the leads (increasing $P$),
making the conductance peak higher and narrower, and adjusts  
the dot-defining gate voltages to again maximize the conductance. 
Repeatedly doing this should give the symmetry point with increasing 
accuracy, until one reaches the limit imposed by
inherent asymmetries (disorder, etc).


{\bf Numerical simulations}.
For the above maximisation 
we took $W_{\rm tb} = L$ and only 12 lead modes.
This calls into question two assumptions in the theory.
Firstly, we can no longer assume that
paths in the dot will be well randomized
between collisions with the barrier, since $\tau'_{\rm D} \sim 3.6\tau_0$.
Secondly, we may not be able to neglect other interference effects 
(weak-localization, etc),
since  $\langle G \rangle$ is at most a few $G_0$.
Thus to verify that the effect is as expected 
in such a parameter regime,
we numerically simulated a stadium billiard containing a barrier
with $T_{\rm tb}=1.48 \times 10^{-3}$,
see Fig.~\ref{Fig:numerics}.
We use the recursive Green's function technique
\cite{papmio} working in real space 
for the direction of current propagation (cut into multiple slices) 
and in mode space for the transverse direction. 
Magnetic fields are in a Landau gauge where the 
vector potential is oriented in the transverse direction \cite{gvr}. 
The number of longitudinal slices and transverse modes were
increased until the results converged.  The data shown here 
are for 836 longitudinal slices 
(200 of which are in the outer leads) and 200 transverse modes.
We mimic thermal smearing, at a temperature
of 23\,mK, by averaging over 44 energies 
uniformly distributed over an interval of 0.02\,meV around the 
Fermi energy of 9.02\,meV. We use the effective mass in
GaAs of 0.067\,$m_0$. 
The simulation (data points in Fig.~\ref{Fig:numerics})
clearly shows that the effect exists in this regime.
Indeed, despite the assumptions in its derivation,
the theory (solid curve) agrees surprisingly well with the
numerical data.

{\bf Concluding comment.}
The conductance peak is {\it not} destroyed 
by bias voltages or temperatures greater 
than $\hbar/\tau_{\rm D}$, because the mirror symmetry 
is present at all energies and not just at the chemical potential 
(unlike the electron-hole symmetry for reflectionless tunneling into a 
superconductor).
Large biases or temperatures should still be avoided, as they
increase the decoherence.
We thank M.~Houzet and P.~Brouwer for discussions.

\bibliographystyle{apsrev}

\begin{thebibliography}{9}
\bibitem{footnote:sign}
It applies at all $\theta$ and all energies for 
barriers with smooth cross-sections (WKB analysis) or rectangular
cross-sections.
\bibitem{footnote:kappa}
In Fig.~\ref{Fig:numerics}(a) we choose 
$\eta=0.34$ 
to best fit the data.
\end{thebibliography}

\newpage
\appendix
\section{Appendix (only on arXiv version)}

{\bf Comment on the conductance ratio.}
It is instructive to consider particular limits of 
Eqs. (\ref{eq:Gsym},\ref{eq:Gasym}).
At the symmetry point we observe that
the more times a path returns to the barrier, the more transparent
the interference makes the barrier.  Thus if the path takes an infinite time to
escape the double dot ($P=1$), then the barrier becomes completely transparent.
However this does not generate a large conductance peak,
since for $P \to 1$ the probability to go from the left lead to the right lead
is a half for any finite barrier transparency, thus
$\langle G_{\rm sym} \rangle/\langle G_{\rm asym} \rangle \to 1$.
This is the reason for the maximum conductance peak occurring 
when $P$ is slightly less than one
(as  visible in Fig.~\ref{Fig:cond-ratio}).
The opposite limit is $P \to 0$, then both  $\langle G_{\rm sym} \rangle$ and 
$\langle G_{\rm asym} \rangle$ reduce to the conductance of the barrier alone.
Since no path hits the barrier more than once, there can be no
interference induced enhancement of tunneling.

{\bf Comment on fitting $B$-field dependence.}
To make the theory quantitative (for comparison with the numerical simulations)
we assumed that
the area enclosed by each straight path segment 
from one point on the boundary of the left dot to another is uncorrelated 
with the next.  We define this directed area, $a$, as that of the triangle 
made by the two ends of the path segment
and the mid-point of the barrier.  Assuming 
$(eBa/\hbar) \ll 1$ we have  
$\Gamma_B= 2\kappa(eB{\cal A}/h)^2/\tau_0$,
where the system-specific parameter 
$\kappa={\rm var}(a/{\cal A})$. 
If $a$ were uniformly distributed over 
the range from $-{\cal A}/2$ to ${\cal A}/2$, we would get 
$\kappa=1/12$.  In contrast, if the distribution were
strongly peaked at $-{\cal A}/2$ and ${\cal A}/2$, we would get
$\kappa$ as big as $1/2$.
We believe only a ray-tracing simulation of the cavity
would yield an accurate value for $\kappa$,
thus for the theory curve in Fig.~\ref{Fig:numerics}a
we treat $\kappa$ as a fitting parameter. 
The Lorentzian width 
of $40\,\mu{\rm T}$ corresponds to $\kappa= 0.17$;
which is within the range estimated above.

{\bf Comment on fitting barrier-position dependence.} 
In the case shown in Fig.~\ref{Fig:numerics}b, 
the mirror symmetry is only broken at the point where a
path segment begins or end at the tunnel barrier. Thus a path segment
only acquires a phase-difference from its mirror image 
at the places where it touches the barrier.
Paths acquire more phase in the left dot than the right dot 
(since the barrier is moved to the right).
This phase difference has a very different form from that induced 
by a $B$-field (where the phase difference grows with the time
a path spends in one of the dots).
Taking this into account, we get the solid curve in 
Fig.~\ref{Fig:numerics}(b) without any fitting parameters.

\end{document}